\documentclass[fleqn,twoside]{article}
\usepackage{espcrc2}

\usepackage[figuresright]{rotating}

\newcommand{\AmS}{{\protect\the\textfont2
  A\kern-.1667em\lower.5ex\hbox{M}\kern-.125emS}}

\title{Polyakov Loops, Z(N) Symmetry, and Sine-Law Scaling}
       
\author{Peter N. Meisinger\address[WU]{Dept. of Physics, Washington University, St. Louis, MO 63130 USA}
        and
        Michael C. Ogilvie\addressmark[WU]\thanks{MCO acknowledges financial support from the U.\ S.\ Department of Energy}}
       
\begin{document}

\begin{abstract}
We construct an effective action for Polyakov loops using the
eigenvalues of the Polyakov loops as the fundamental variables. We assume $%
Z(N)$ symmetry in the confined phase, a finite difference in energy
densities between the confined and deconfined phases as $T\rightarrow 0$,
and a smooth connection to perturbation theory for large $T$. The
low-temperature phase consists of $N-1$ independent fields fluctuating around an
explicitly $Z(N)$ symmetric background. In the low-temperature phase, the
effective action yields non-zero string tensions for all representations
with non-trivial $N$-ality. Mixing occurs naturally between representations
of the same $N$-ality. Sine-law scaling emerges as a special case,
associated with nearest-neighbor interactions between Polyakov loop
eigenvalues.\vspace{1pc}
\end{abstract}

\maketitle

Effective actions for the Polyakov loop are directly relevant to the phase
diagram and equation of state for QCD and related theories
\cite{Pisarski:2000eq,Dumitru:2001xa,Meisinger:2001cq,Dumitru:2003hp,Meisinger:2003id}. 
Furthermore,
there may be a close relationship between the correct effective action and
the underlying mechanisms of confinement
\cite{Meisinger:1997jt,Meisinger:2002ji,Kogan:2002yr,Mocsy:2003tr,Mocsy:2003un}.
We construct a
general effective action for the Polyakov loop
which describes both the
confined and deconfined phases of $SU(N)$ pure gauge theories at finite
temperature, making as few assumptions as
possible\cite{Meisinger:2004new}.

It is widely held that the asymptotic string tension, obtained
from Wilson loops or Polyakov loop two-point functions,
depends only on the $N$%
-ality of the representation. Lattice data from simulations of four
dimension $SU(3)$ gauge theory do not yet show this behavior \cite{Deldar:1999vi,Bali:2000un}. 
Instead,
the string tension $\sigma _{R}$ associated with a representation $R$ scales
approximately as 
$\sigma _{R}=(C_R/C_F)\sigma_F$
where $C_{R}$ is the quadratic Casimir invariant for the representation $R$. 
We will
refer to this behavior as Casimir scaling,
and use the term $Z(N)$ scaling for
the asymptotic behavior
$ \sigma_R=(k(N-k))/(N-1))\sigma _{F} $
where $k$ is the $N$-ality of the representation $R$. Another possible scaling law for the string tensions is sine-law
scaling
$ \sigma_R=(\sin (\pi k/N)/(\sin (\pi/N))\sigma _{F} $
which has been shown to occur in softly broken $\emph{N}=2$ super Yang-Mills
theories\cite{Douglas:1995nw}
and in MQCD\cite{Hanany:1997hr}.

In a gauge in which $A_{0}$ is time independent and diagonal, we may write $%
P $ in the fundamental representation as 
$ P_{jk}=\exp \left( i\theta _{j}\right) \delta _{jk} $
so all information about $P$ is contained in the $N-1$ independent eigenvalues.
In the confined phase, we look for a $Z(N)$-symmetric field configuration,
which implies $zP_{0}=gP_{0}g^{+}$.
This in turn implies $Tr_{R}P_{0}=0$ for all representations $R$
with non-zero $N$-ality, which means that all representations with non-zero $%
N$-ality are confined. The diagonal form of $P_{0}$ is 
$ d=w\,\,diag\left[ z,z^{2},..,z^{N}=1\right] $
where $z$ is henceforth $\exp \left( 2\pi i/N\right) $, the generator of $%
Z(N)$. The phase $w$ ensures that $d$ has determinant $1$, and is given by $%
w=\exp \left[ -\left( N+1\right) \pi i/N\right] $. 
The corresponding eigenvalues are
$ \theta _{j}^{0}==\frac{\pi }{N}\left( 2j-N-1\right)$, which is
uniform spacing
around the unit circle.


In the low-temperature, confining phase, we
consider small fluctuations about $P_{0}$, defining $\theta _{j}=\theta
_{j}^{0}+\delta \theta _{j}$. 
Although this approximation may not be {\it a priori} valid,
the assumption that fluctuations are small
can be justified in the large-$N$ limit. 
In the case of non-zero $N$-ality,
we have for small fluctuations that
$Tr_{F}P^{k}=ikw^{k}\phi _{k}$,
where $\phi_k$ is the discrete Fourier transform
of $\delta \theta _{j}$.
Operators with the same $N$-ality generally give inequivalent expressions
when written in terms of the the $\phi $ variables. For example, $%
Tr_{F}P^{k}\propto \phi _{k}$ and $\left( Tr_{F}P\right) ^{k}\propto
\left( \phi _{1}\right) ^{k}$.

A sufficiently general form of the action at all temperatures has the form 
\begin{equation}
S_{eff}=\beta \int d^{3}x\left[ \kappa T^{2}Tr_{F}\,\left( \nabla \theta
\right) ^{2}+V(\theta )\right] 
\end{equation}
where $\kappa $ is a temperature dependent correction to the kinetic term,
and $V$ is a class function only of the adjoint eigenvalues $\theta _{j}-\theta
_{k}$. We assume that there is a finite free energy density difference associated
with different values of $P$ as $T\rightarrow 0$. Because the eigenvalues
are dimensionless, this requires terms in the potential with coefficients
proportional to $\left( mass\right) ^{4}\,$as $T\rightarrow 0$.
The quadratic terms in the effective action are diagonalized when
expressed in terms of the modes $\phi_n$; we write this as
\begin{equation}
\frac{\kappa T^{2}}{N}%
\sum_{n=1}^{N-1}\left( \nabla \phi _{n}\right) \left( \nabla \phi
_{N-n}\right)+\sum_{n=1}^{N-1}M_{n}^{4}\phi
_{n}\phi _{N-n} \rm{.}
\end{equation}
If we write the
higher-order terms of $S_{eff}$ in terms of the $\phi _{n}$,
each interaction will manifestly respect global conservation of $N$-ality. 
For example,
in $SU(4)$, an interaction of the form $\phi _{1}^{2}\phi _{2}$ is allowed,
but not $\phi _{1}^{2}\phi _{2}^{2}$. These interactions lead to mixing
of operators with identical $N$-ality, and thus scaling based
on $N$-ality rather than representation.

The confining behavior of Polyakov loop two-point functions at low
temperatures, for all representations of non-zero $N$-ality, is natural in
the effective model. If the interactions are neglected, we can calculate the
behavior of Polyakov loop two-point functions at low temperatures from the
quadratic part of $S_{eff}$. We have for large distances that the two-point 
functions fall off as
$\exp \left[ -\sigma _{n}\left| x-y\right|/T \right]$
where $\sigma _{n}\left( T\right) =\sqrt{NM_{n}^{4}(T)/\kappa (T)}$ is
identified as the string tension for the $n$'th mode at temperature $T$.
The zero-temperature string
tension is given at tree level by 
\begin{equation}
\sigma _{n}^{2}\left( 0\right) =NM_{n}^{4}(0)/\kappa (0) 
\end{equation}

\vspace{1pt}Sine-law scaling arises naturally from a nearest-neighbor
interaction in the space of Polyakov loop eigenvalues. Consider the class of
potentials with pairwise interactions between the eigenvalues 
$V_{2}=\sum_{j,k}v\left( \theta _{j}-\theta _{k}\right)$. 
as obtained, for example, 
by two-loop perturbation theory\cite{KorthalsAltes:1993ca}.
An elementary calculation shows that at tree level 
\begin{equation}
\sigma _{n}=\sqrt{\frac{2}{\kappa }\sum_{j=0}^{N-1}v^{\left( 2\right)
}\left( \frac{2\pi j}{N}\right) \sin ^{2}\left( \frac{\pi nj}{N}\right) } 
\label{eq:dispersion}
\end{equation}
where $v^{(2)}$ is the second derivative of $v$.
If the sum is dominated by the $j=1$ and $j=N-1\,$terms,
representing a nearest-neighbor interaction in the space of eigenvalues,
then we recover sine-law scaling
$\sigma _{n} \propto \sin (\pi n/N) \rm{.}$
$Z(N)$ scaling can be obtained by a very small admixture of other components of $%
v^{\left( 2\right) }$, as we show below.

The string tension associated with different $N$-alities has been measured
in $d=3$ and $4$ dimensions for $N=4$ and $6$, and in $d=4$ for $N=8$
\cite{Lucini:2001nv,Lucini:2002wg,DelDebbio:2001sj,DelDebbio:2003tk,Lucini:2004my}.
We examine the simulation data by inverting equation 
(\ref{eq:dispersion}) above to give a
measure of the relative strength of the couplings $v^{\left( 2\right)
}\left( 2\pi j/N\right) $. We normalize the result of this inversion such
that the sum of the independent couplings adds to one, and the results are
shown in Table \ref{Table1}. 
Sine-law scaling corresponds to a value of $1\,$for $j=1$, 
and $0$ for the other $\left[ N/2\right] -1$ independent couplings. For $%
Z(N)$ scaling, the large-$N$ limit gives $v^{\left( 2\right) }\left( 2\pi
j/N\right) \propto 1/j^{4}$, yielding the result shown in the table. Note
that the difference between sine-law scaling and $Z(N)$ scaling remains
small but finite, even as $N$ goes to infinity. The three-dimensional
simulation results clearly favor $Z(N)$ scaling, but smaller error bars 
will be needed to differentiate between
sine-law and $Z(N)$ scaling in four dimensions.

\begin{table*}
\caption{{\label{Table1}}Relative strength of couplings $v^{\left( 2\right)}\left( 2\pi j/N\right) $}
\begin{tabular}{|l||l|l|l|l|l|}
\hline
& $d$ & $j=1$ & $j=2$ & $j=3$ & $j=4$ \\ \hline\hline
$SU(4)$\cite{Lucini:2001nv,Lucini:2002wg} & 3 & 0.957(6) & 0.043(2) &  &  \\ \hline
$SU(4)$\cite{Lucini:2004my} & 4 & 0.968(16) & 0.032(7) &  &  \\ \hline
$SU(4)$\cite{DelDebbio:2001sj,DelDebbio:2003tk} & 4 & 0.992(25) & 0.008(5) &  & \\ \hline
$SU(4)\,\,Z(N)$ & any & 0.9412 & 0.0588 &  &  \\ \hline
$SU(6)$\cite{Lucini:2001nv,Lucini:2002wg} & 3 & 0.930(5) & 0.065(8) & 0.004(5) &  \\ \hline
$SU(6)$\cite{Lucini:2004my} & 4 & 0.960(16) & 0.045(18) & -0.005(12) &  \\ \hline
$SU(6)$\cite{DelDebbio:2001sj,DelDebbio:2003tk} & 4 & 0.996(40) & 0.0003(218) & 0.004(26) &  \\ \hline
$SU(6)\,\,Z(N)$ & any & 0.9266 & 0.0618 & 0.0116 & \\ \hline
$SU(8)$\cite{Lucini:2004my} & 4 & 1.028(22) & -0.067(24) & 0.047(32) & -0.009(22) \\ \hline
$SU(8)\,\,Z(N)$ & any & 0.9249 & 0.0583 & 0.0130 & 0.0037 \\ \hline
$Z(N)$ $N\rightarrow \infty $ & any & 0.9239 & 0.0577 & 0.0114 & 0.0036 \\ \hline
sine Law & any & 1 & 0 & 0 & 0 \\ \hline
\end{tabular}
\end{table*}

In previous work on phenomenological models
of the gluon equation of state\cite{Meisinger:2001cq,Meisinger:2003id}, 
we considered models with free energy
$f = V(\theta ) - p_g(\theta)$
where $p_g$ is the pressure of a gauge boson gas moving 
in a constant background Polyakov
loop, and $V$ is a phenomenologically chosen potential favoring 
confinement at low temperature. We studied two
potentials which reproduce $SU(3)$ thermodynamics well,
and correctly describe the deconfinining phase transition
for all $N$. The first potential is a quadratic function of the eigenvalues
and leads to
$\sigma _{k}^{A}=\sigma _{1}$ for every 
$N$-ality. The other potential is the logarithm of Haar measure,
and leads to 
$\sigma _{k}^{B}\propto \sqrt{k\left( N-k\right) }$,
which one might call ''square root of Z(N) scaling''. Although these two
models are not consistent with the lattice simulation data for $\sigma _{k}$
for $N>3$, they remain viable phenomenological forms for $N=2$ and $3$. It
is interesting to note that there is a well-studied potential which gives $%
Z(N)$ scaling\cite{Calogero:1978}. It is the integrable 
Calogero-Sutherland-Moser potential 
\begin{equation}
V_{Z}\left( \theta \right) =\sum_{j\neq k}\frac{\lambda }{\sin ^{2}\left( 
\frac{\theta _{j}-\theta _{k}}{2}\right) } 
\end{equation}
which has been associated with
two-dimensional gauge theory\cite{Minahan:1993np,Polychronakos:1999sx}.

\end{document}